# Development of 3D-DDTC pixel detectors for the ATLAS upgrade

Gian-Franco Dalla Betta[a,*], Maurizio Boscardin[b], Giovanni Darbo[c], Claudia Gemme[c], Alessandro La Rosa[d], Heinz Pernegger[d], Claudio Piemonte[b], Marco Povoli[a], Sabina Ronchin[b], Andrea Zoboli[a], Nicola Zorzi[b]

[a] INFN, Sezione di Padova (Gruppo Collegato di Trento), and DISI, Università di Trento, Via Sommarive 14, 38123 Povo di Trento, Italy

[b] Fondazione Bruno Kessler (FBK-irst), Via Sommarive 18, 38123 Povo di Trento, Italy

[c] INFN, Sezione di Genova, Via Dodecaneso 33, 16146 Genova, Italy

[d] CERN – PH, CH-1211 Geneve 23, Switzerland

## Abstract

We report on the development of n-on-p, 3D Double-Side Double Type Column (3D-DDTC) pixel detectors fabricated at FBK-irst (Trento, Italy) and oriented to the ATLAS upgrade. The considered fabrication technology is simpler than that required for full 3D detectors with active edge, but the detector efficiency and radiation hardness critically depend on the columnar electrode overlap and should be carefully evaluated. The first assemblies of these sensors (featuring 2, 3, or 4 columns per pixel) with the ATLAS FEI3 read-out chip have been tested in laboratory. Selected results from the electrical and functional characterization with radioactive sources are here discussed.

---

[*] Corresponding author. Tel.: +39-0461883904; fax: +39-0461882093; e-mail: dallabe@disi.unitn.it

# 1. Introduction

Silicon 3D detectors consist of an array of columnar electrodes of both doping types, oriented perpendicularly to the wafer surface and penetrating entirely through the substrate [1]. Provided the distance between columns is properly designed, this architecture allows for low depletion voltage and fast charge collection times, while keeping the detector active thickness unaltered. These properties can effectively counteract charge trapping effects due to high levels of radiation in HEP experiments. As a result, 3D detectors are expected to be more radiation tolerant than planar detectors and are emerging as one of the most promising technologies for innermost layers of tracking at the foreseen upgrades of the Large Hadron Collider. As an additional feature, 3D technology is suitable for manufacturing detectors with "active edge" [2], i.e., terminated with heavily doped trenches, where the insensitive edge region can be reduced to a few μm, to be compared to a few hundreds of μm for standard planar detectors. This option can facilitate the overall detector layout and reduce the material budget, since no sensor overlap is needed within the same layer.

Full-3D detectors with active edges fabricated at the Stanford Nano Fabrication Facility (in the following referred to as standard 3D detectors), are the state-of-the-art in this field and have already proved to yield high good performance and high radiation tolerance. As an example, infrared laser tests on these sensors irradiated with neutrons and read-out with a fast transimpedance amplifier have shown a signal efficiency as high as 66% after a fluence of $8.8 \times 10^{15}$ 1-MeV $n_{eq.}/cm^2$ [3]. Moreover, standard 3D pixel detectors bump-bonded to the ATLAS FEI3 front-end chip [4] have been measured in a 100 GeV pion beam at CERN SPS [5]. A hit efficiency of 99.9%±0.1% has been reported in case of 15° track inclination, which largely suppresses efficiency losses due to the electrodes, at the expense of larger charge sharing between adjacent pixels (i.e., larger cluster sizes).

Besides standard 3D detectors, other modified 3D detectors have so far been proposed, among them 3D-STC (Single Type Column [6]) and 3D-DDTC (Double-side Double Type Column) detectors, aiming at a simplification of the manufacturing technology in view of volume productions. In particular, the 3D-DDTC approach (in two slightly different versions) is being pursued by CNM-IMB (Barcelona, Spain [7]) and by FBK-irst (Trento, Italy [8]) with encouraging results.

3D detector technologies have been developed at FBK-irst in the framework of a collaboration with INFN since 2004. In the past few years, this activity has been mainly focused on sensors oriented to the upgrade of the ATLAS Pixel Detector, and the 3D-DDTC approach is currently considered a possible alternative to standard 3D design in the framework of the CERN 3D-ATLAS Sensor Collaboration [9].

Using 3D-DDTC technology, we have fabricated at FBK-irst two batches of detectors on p-type substrates. In these detectors, the columns are not passing through the entire wafer thickness. The wafer layout is mainly oriented to pixel detectors compatible with existing ATLAS read-out chips. A few samples of 3D ATLAS pixel detectors with 2, 3, or 4 columns per pixel were bump bonded to the ATLAS FEI3 chips at SELEX SI (Rome, Italy) [10] and the assemblies have been tested electrically and with radioactive sources.

In this paper we report on these detectors, covering design and technological aspects and selected results from the experimental characterization.

## 2. Sensor description

Sensors have been fabricated at FBK-irst MT-Lab. (Trento, Italy) on Float Zone, p-type, high-resistivity silicon wafers using the 3D-DDTC technology [8]. Columnar electrodes of both doping types are etched from both wafer sides (junction columns from the front side, Ohmic columns from the back side), and stopping at a short distance (ideally not exceeding a few tens of micrometers) from the opposite surface. Junction columns ($n^+$) are read-out columns and are arranged in the pixel configuration connecting them by a surface $n^+$ diffusion and a metal strip. Ohmic columns ($p^+$) are all connected together by a uniform surface $p^+$ diffusion and metallization on the back side. All columns have a nominal diameter of 10 μm and are not filled with poly-Si. As an example, Fig.1a shows the layout of two adjacent pixels, whereas a schematic cross-section of the sensors is shown in Fig.1b. Surface insulation between $n^+$ pixels is achieved by combined p-spray/p-stop implants [11]. The fabrication technology is similar to that described in [8] for 3D-DDTC detectors made on n-type substrates, except for: (i) the substrate type, (ii) the inverted doping of the columns and related surface regions, and (iii) the additional steps for p-spray/p-stop implantations on the front side.

The peculiar shape of present ATLAS pixels (50 μm × 400 μm) lends itself to different choices in terms of number of columns per pixel, and, accordingly, of pitch between the columns. In our design, we have implemented several layout options. Among them, those featuring two (2E), three (3E), and four (4E) junction columns per pixel, which are schematically represented in Fig.2, will be considered in this paper. Besides ATLAS pixel detectors, the wafer layout contains CMS pixel detectors, strip detectors and test structures (both planar and 3D).

Two batches of detectors have been fabricated. In the first one (3D-DTC-2), completed in July 2008, the DRIE steps had to be performed as an external service, because the equipment was not available at FBK yet. This significantly delayed the fabrication, caused major yield problems, and also limited the maximum etching depth achievable for the columns. When the DRIE equipment (Adixen AMS200) became available at FBK, a fabrication recycle (3D-DTC-2B) could be processed entirely in house and was completed in April 2009. Table 1 summarizes the main geometrical parameters of detectors from the two batches. The substrate thickness values have been extracted from the C-V curves of planar test diodes. The column thickness values have been extracted from the C-V curves of 3D test structures. As can be seen, deeper junction columns and a larger overlap between columns of different doping type have been achieved in the second batch, from which better charge collection characteristics are therefore expected [8].

## 3. Electrical characterization and TCAD simulations

Test structures from the two batches have been extensively measured on wafer before proceeding with functional tests. Measurements were performed at room temperature and under dark conditions by using a probe-station to contact the devices. Results from the electrical characterization of devices belonging to the first batch are reported in [12]. For the sake of clarity, they are summarized in Table 2 along with results relevant to the second batch. The substrate doping concentration was extracted from the C-V curves of planar test diodes, whereas all other data were extracted from 3D test diodes having 80 μm pitch between columns of the same doping type, corresponding to ~56 μm pitch between columns of opposite doping type (i.e., comparable to ATLAS pixel detectors of type 4E). The value of those parameters (depletion voltage, capacitance vs backplane) depending on the substrate doping concentration and column thickness are slightly different in the two batches, as expected. In particular, the full depletion voltage is sensibly lower in the second batch. On the contrary, the values of the technology dependent parameters, such as leakage current and breakdown voltage, are very similar, evidence of the good reproducibility of FBK process. As for the breakdown voltage, it is normally larger than 70 V. The "intrinsic" value is indeed higher than 100 V and is determined by the $n^+$/p-spray junction at the top surface. Due to some defects, lower values are sometimes observed.

In [8], TCAD simulation results addressing the charge collection properties of 3D-DDTC detectors are reported. Using results extracted from the measurements, new TCAD simulations have been performed, incorporating exact geometrical and process parameters relevant to detectors from the first batch. As an example, Fig. 3 shows the simulated current signal in a detector reverse biased at 16 V in response to a minimum ionizing particle impinging perpendicularly to the surface and close to the Ohmic column. The evolution of the electron density distribution is shown in the insets at several time instants (t=0, 0.5 ns, 1.0 ns, and 1.5 ns). The current signal exhibits a very fast component owing to the rapid charge collection from the high-field region (i.e., where columns overlap), and a slower tail due to diffusion from regions with low electric field (it should be stressed that full depletion is not easy to achieve at the bottom of the wafer close to the Ohmic columns). Nevertheless, this result confirms that, in spite of non optimized junction column depth, the full charge collection process in these detectors requires just a few ns, so that good charge collection efficiency is expected even with fast read-out electronics, like that used for ATLAS pixels. More details on simulations can be found in [13], along with experimental results from transient current characterization of 3D diodes stimulated with IR laser pulses. Further confirmation of the performance enhancement of these detectors with respect to the previous 3D-DDTC detectors made on n-type substrates [14] came from results of functional tests with laser and beta source setup performed on strip detectors from batch 3D-DTC-2 connected to the ATLAS SCT Endcap electronics (ABCD3TA chip) [15], for which a better spatial uniformity of the signal and no sign of ballistic deficit have been observed.

## 4. Single-chip assembly

A few pixel sensors from both batches have been connected via bump bonding and flip chip to the ATLAS FEI3 read-out chip [4]. The bump bonding process is based on Indium and it has been carried out at SELEX SI [10]. The FEI3 chip consists of 2880 cells of 50μm × 400μm size arranged in a matrix of 18 × 160, matching the geometrical characteristics of the sensor. In each cell, the corresponding pixel charge signal is amplified and compared to a programmable threshold by a discriminator. The digital readout provides information on the hit pixel address, the hit time stamp and the digitized amplitude, expressed in terms of Time-over-Threshold (ToT – length of discriminator signal). The ToT of a hit is determined by the width of the injected pulse and depends on: the deposited charge, the discriminator threshold and the feedback current. To measure the charge of a hit the discriminator output pulse is recorded in units of the 40 MHz crossing clock [4].

## 5. Functional characterization

The experimental setup used for the characterization of the detector is based on the TurboDAQ setup, which has been developed at LBNL [16] and used to perform automated electrical test of ATLAS Pixel Detector Modules during the production phase. It runs under Windows and is based on National Instruments LabWindows development suite.
Figure 4 shows a snapshot of CERN ATLAS Pixel setup, which is also equipped with a climate chamber that can be operated in a range between -25°C and 100°C, where detectors under test are housed. All measurements here reported have been done at 20°C with a relative humidity of 12%. The performance of 3D-DDTC sensors have been studied by measuring leakage currents, threshold and noise, and response to γ ($Am^{241}$ and $Cd^{109}$) and β ($Sr^{90}$) radioactive sources.

*5.1 Leakage current measurements.*

Figure 5 provides an overview of the I-V curves of all 3D-DDTC sensors available from the first batch. This kind of test has been also useful to check for sensor damages after different stages on the assembly. From Fig. 5 it can be seen that in most detectors breakdown takes place at voltages of about 70 V, in good agreement with measurements on test structures. Two samples show breakdown at lower voltages, probably due to some damage during the assembly. The leakage current below breakdown is quite large as compared to predictions based on the measurements performed on test diodes, but it still remains at an acceptable level (~100 pA/pixel). As an example, Fig. 6 shows the pixel leakage current distribution in a 2E sample.

As for the sensors from the second batch, a sharp increase of the current up to the compliance set at 10 µA has been observed already at low voltage (10 – 12 V). The four sensors so far tested have all shown the same behaviour (only 3E sensors have been initially considered for the second batch in view of their use in a beam test at CERN). Such an early breakdown is certainly related to the presence of local defects. As it can be seen from Fig. 7, showing a map of the pixel leakage current at different voltages, a few pixels start exhibiting high leakage current as the voltage is increased beyond 10 V. It should ne noted that it was not possible to make direct electrical tests on pixel detectors before assembling them with the FEI3 chip. On the other hand, no evidence of systematic early breakdown could be observed in strip detectors, which have a comparable total size (in the order of 1 cm$^2$), so that the defect density is not believed to be so high to justify this behaviour in each sensor. The reason for this behaviour is likely to be ascribed to some damage occurred during the assembly. As a matter of fact, also planar test structures belonging to the same wafer have shown a degradation of their leakage current (up to a factor of 20) after bump bonding. This point has not been understood and is still under investigation.

*5.2    Threshold and noise measurements*

This test has been performed to measure the threshold and noise of each pixel, where only pixels with a charge deposit above the threshold are taken into account for readout by the front-end electronics. Signals are induced in each pixel by means of on-chip charge injectors, and the number of collected hits versus the injected charge is recorded. In an ideal case a step function with an immediate transition of the detection efficiency from 0 to 100 % at the threshold would be expected. In the real case, because of pixel noise, some injected charges below the threshold are detected and some injected charges above the threshold are not detected. The error function, a convolution of the ideal step function with the Gaussian pixel noise distribution, describes the discriminator output. This function, the so-called S-curve, is fitted to the threshold scan result of each pixel. The 50% efficiency on the S-curve defines the threshold value of a pixel. The noise of a pixel is inversely proportional to the steepness of the transition from no detected hits to full efficiency [17]. Scans are repeated to reduce the threshold dispersion by adjusting a DAC-parameter individually for each channel [4,18]. The on-chip injection circuits are also used to calibrate the ToT response of a signal into a charge value. The standard tuning aims at a ToT of 60 units for a charge of 20,000 e (which is the most probable charge deposit of a minimum ionizing particle in a silicon sensor of ~250 μm thickness [19]). Given a standard threshold of 3,200 e, this corresponds to a charge of about 250 – 350 e per ToT unit.

The results of the Threshold-Scan (performed at a bias voltage of 35 V) on sensors from the first batch are summarized in Table 3. As an example, Figs. 8 and 9 show a threshold measurement and a noise distribution for a 2E sensor.

The effect of the bias voltage on the noise has been also studied. Figure 10 shows the noise vs voltage curves of all "good" sensors from the first batch (i.e., those not affected by early breakdown problems). As expected, the noise decreases with bias, since the main contribution comes from the pixel capacitance, which is also decreasing with bias. Three different noise levels are eventually distinguished for the three types of sensors, in good agreement with their different capacitance values, which could be measured independently using a strip-like test structure featuring the same column configurations as the ATLAS pixels, and yielding: $C_{2E}$ = 250 fF, $C_{3E}$ =310 fF, $C_{4E}$=370 fF. The noise values are lower than those reported in [3] for standard 3D sensors coupled to the FEI3 chip, as a result of a lower column overlap, and only slightly higher than those reported for planar pixel sensors (160 e, [19]).

The noise behaviour of sensors from the second batch is very similar at low voltage, with values of about 240 e at 10 V, whereas it is obviously degraded as the voltage is increased because of early breakdown. A good correlation was indeed found between leakage current and noise, as shown in the maps of Fig. 11. As a matter of fact, the regions of high noise clearly develop around the leaky pixel seeds (see also Fig.7).

## 5.3 Measurements with radioactive sources

$Am^{241}$ and $Cd^{109}$ γ sources have been chosen to calibrate the detectors. The source tests have been also used to identify dead pixels that do not show any signal because they are disconnected, merged, defective or badly tuned. $Am^{241}$ ($Cd^{109}$) γ source emits 60 keV (22 keV) photons which can convert anywhere in the bulk to a 60 keV (22 keV) primary electron. If ionization takes place in the substrate region where columns overlap, a signal of 16.5 ke (6.1 ke) is expected. On the other hand, if a photon converts in a high doping region or close to the surfaces, a fraction of the charge could be lost. Thus, in the charge distribution, a high-end pick at 16.5ke (6.1ke) and a tail towards smaller values are expected.

Figure 12 shows spectra acquired with a 2E sensor for $Am^{241}$ and $Cd^{109}$, respectively, as reconstructed from the ToT-reading with the calibration shown in Fig. 13. Data refer to a bias voltage of 35 V. In both cases, the position of the main peaks agrees with expectations within the uncertainty due to the calibration process, which was estimated to be in the order of 10-15%. Results are also in good agreement with those obtained from measurements on ATLAS Planar Pixel Sensor [19] single-chip module using the same setup and with data already published in [20] for the ATLAS Pixel detector.

Figure 14 shows the pulse height distributions in response to a $Sr^{90}$ β source in a 2E sensor biased at 35 V for cluster size 1 and 2. Charge values have been fitted with Landau functions, which are also shown in the figures. The most probable value (MPV) of the charge is ~14,100 e for cluster size 1 and increases up to ~15,360 for cluster size 2. With cluster size 2, the tail of low charge events clearly visible for cluster size 1 and possibly due to charge sharing disappears.

It should be noted that the front-end electronics has been tuned with 60 ToT at 20,000 e for a planar pixel sensor of ~250 μm, whereas the thickness of 3D sensors is about 200 μm. The charge MPV values measured in 3D sensors with different column configurations are reported in Table 4. No appreciable differences between the three layouts are observed, as indeed expected before irradiation. Table 4 also includes data relevant to a reference planar sensor of 250 μm thickness. As can be seen, the MPV are properly scaling with the sensor thickness, within the already mentioned uncertainties due to the calibration process, so that it can be concluded that charge collection process in these 3D sensors is fully efficient before irradiation.

Measurements with sources are being taken also for the sensors of the second batch, although they can be operated in a narrow bias voltage range (from 4 to 8 V) because of the early breakdown problems.

## Conclusion

We have reported on 3D-DDTC pixel sensors fabricated at FBK-irst. The design and technological characteristics have been reviewed and selected results from the electrical and functional characterization of the first assemblies of these sensors with the ATLAS FEI3 read-out chip have been presented. After the assembly process, some degradation of the electrical characteristics of the sensors, both in terms of leakage current and breakdown voltage, have been observed and are currently being investigated. On the assemblies not affected by early breakdown problems, the noise figures are good (200-240 electrons rms at full depletion), the different values being due to the different capacitance resulting from the different number of columns per pixel and the different distance between columns. Functional tests with γ- and β-sources have shown good performance in terms of charge collection efficiency, although the sensors are still not optimized in terms of column overlap. The discrepancies between the experimental results and the theoretically expected values could be explained taking into account the 10-15% errors due to the calibration process.

A couple of these sensors have been included in a beam test at CERN in May 2009, and preliminary results from data analysis are very encouraging. Radiation hardness tests are under way.

## Acknowledgement

This work has been supported in part by the Provincia Autonoma di Trento and in part by the Italian National Institute for Nuclear Physics (INFN), Projects TREDI (CSN5) and ATLAS (CSN1).

We would like to thank: G. Gariano, A. Rovani ed E. Ruscino (INFN Genova), F. Rivero (University of Torino), and J.-W. Tsung (University of Bonn), for their precious help in system assembly and measurements; R. Beccherle (INFN Genova) for designing bump bonding mask, and S. Di Gioia (Selex SI) for the bump bonding process.

Table 1. Main geometrical parameters of detectors from the two batches (all values are in μm).

| Parameter | Value | |
|---|---|---|
| | 3D-DTC-2 | 3D-DTC-2B |
| Substrate thickness | 200 | 200 |
| Junction column thickness | 100 - 110 | 140 - 170 |
| Ohmic column thickness | 180 – 190 | 180 - 190 |
| Column overlap | 90 - 100 | 110 - 150 |

Table 2. Summary of the electrical parameters extracted from test diodes from the two batches.

| Parameter | Unit | Value | |
|---|---|---|---|
| | | 3D-DTC-2 | 3D-DTC-2B |
| Substrate doping concentration | $cm^{-3}$ | $1 \times 10^{12}$ | $7 \times 10^{11}$ |
| Lateral depletion voltage | V | 3 | 1 - 2 |
| Full depletion voltage | V | 12 | 3 - 4 |
| Capacitance vs backplane | fF/column | 35 | 45 - 50 |
| Leakage current @ Full depletion | pA/column | < 1 | < 1 |
| Breakdown voltage | V | > 70 | > 70 |

Table 3 Summary of the threshold and noise parameters (average and standard deviation) extracted from the pixel sensors from the first batch at a reverse bias voltage of 35 V (all values are in electrons).

| Sensor type | Threshold | Noise |
|---|---|---|
| 2E | $3200 \pm 58.60$ | $202.3 \pm 8.96$ |
| 3E | $3318 \pm 42.02$ | $206.6 \pm 8.29$ |
| 4E | $3284 \pm 41.27$ | $229.8 \pm 9.87$ |

Table 4 Summary of the MPV of the measured charge (average and standard deviation) is response to a $Sr^{90}$ source. Data are relevant to 3D pixel sensors from the first batch biased at 35 V for cluster sizes 1 and 2, and compared to those measured in a reference planar sensor of 250 μm thickness .

| Sensor type | Clu. Size 1 (ke) | Clu. Size 2 (ke) |
|---|---|---|
| 2E | $14.12 \pm 0.03$ | $15.36 \pm 0.05$ |
| 3E | $14.07 \pm 0.03$ | $15.25 \pm 0.02$ |
| 4E | $14.07 \pm 0.03$ | $15.25 \pm 0.03$ |
| Planar | $17.18 \pm 0.18$ | $18.52 \pm 0.06$ |

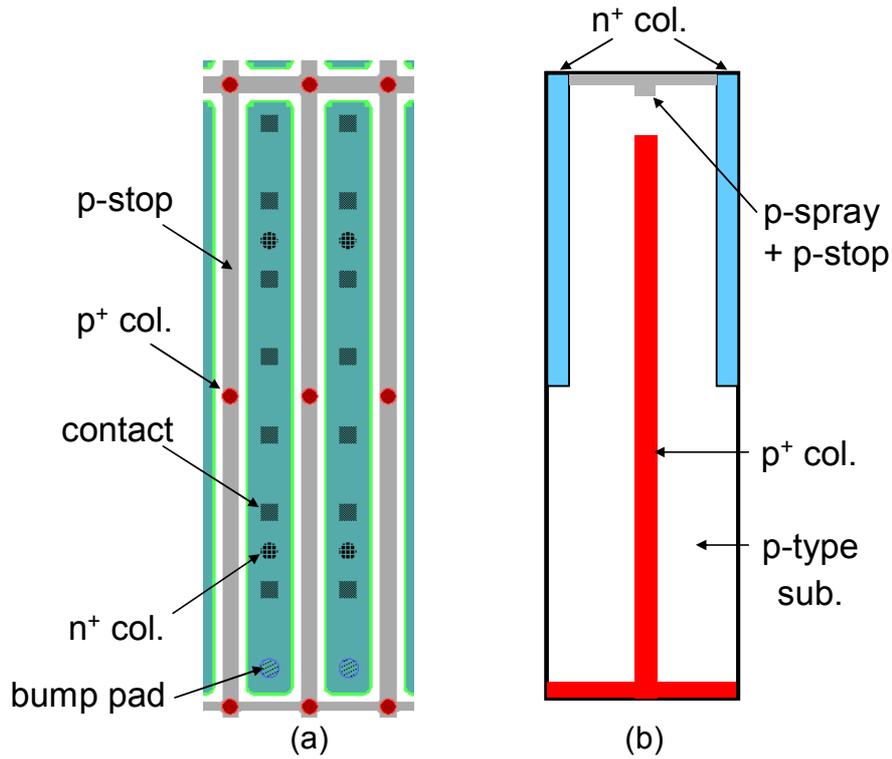

**Fig. 1** (a) Layout detail of ATLAS pixel detectors (2E configuration), and (b) schematic cross-section of the devices along a plane intersecting both types of columns.

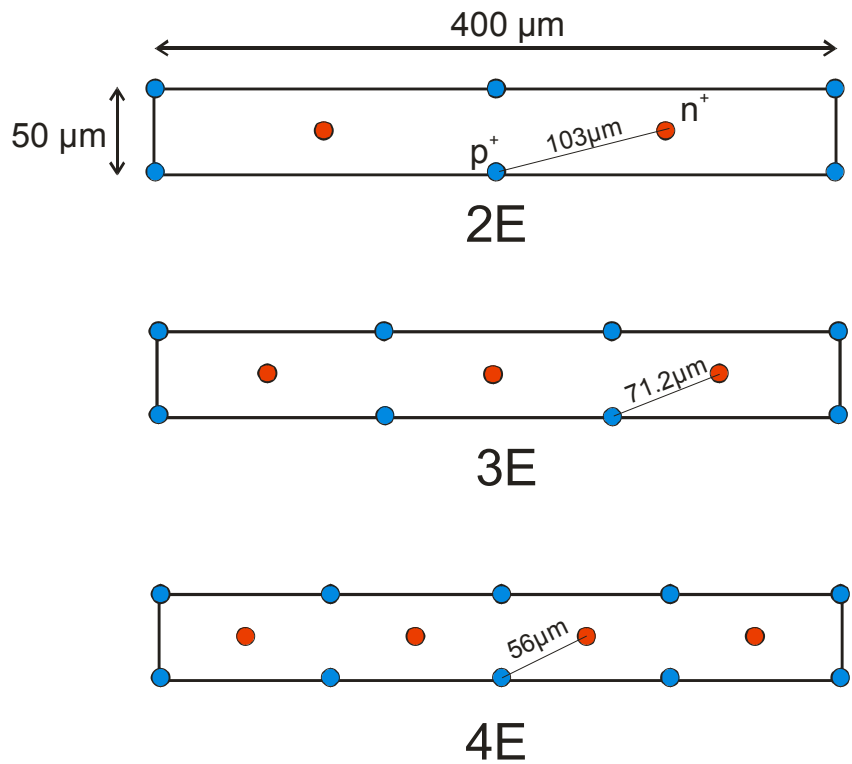

**Fig. 2** Sketch of columnar electrodes in ATLAS pixels of 2E, 3E, and 4E type.

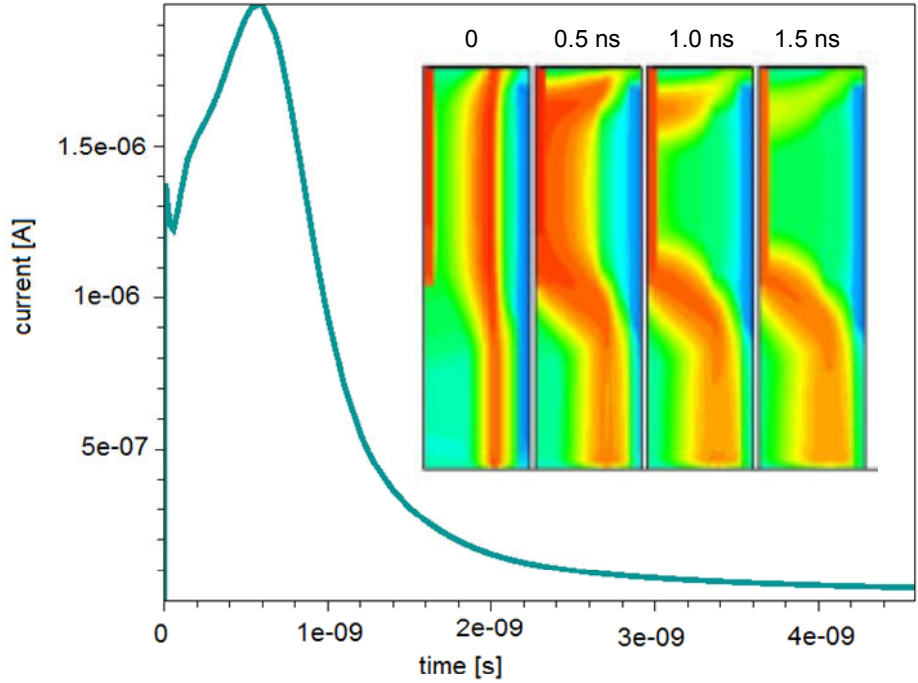

Fig. 3 Simulated transient current signal in a 3D-DDTC diode from batch 3D-DTC-2 reverse biased at 16 V in response to a minimum ionizing particle impinging perpendicularly to the surface and close to the Ohmic column. Evolution of the electron density distribution with time is shown in the inset at instants: t=0, 0.5 ns, 1.0 ns, and 1.5 ns.

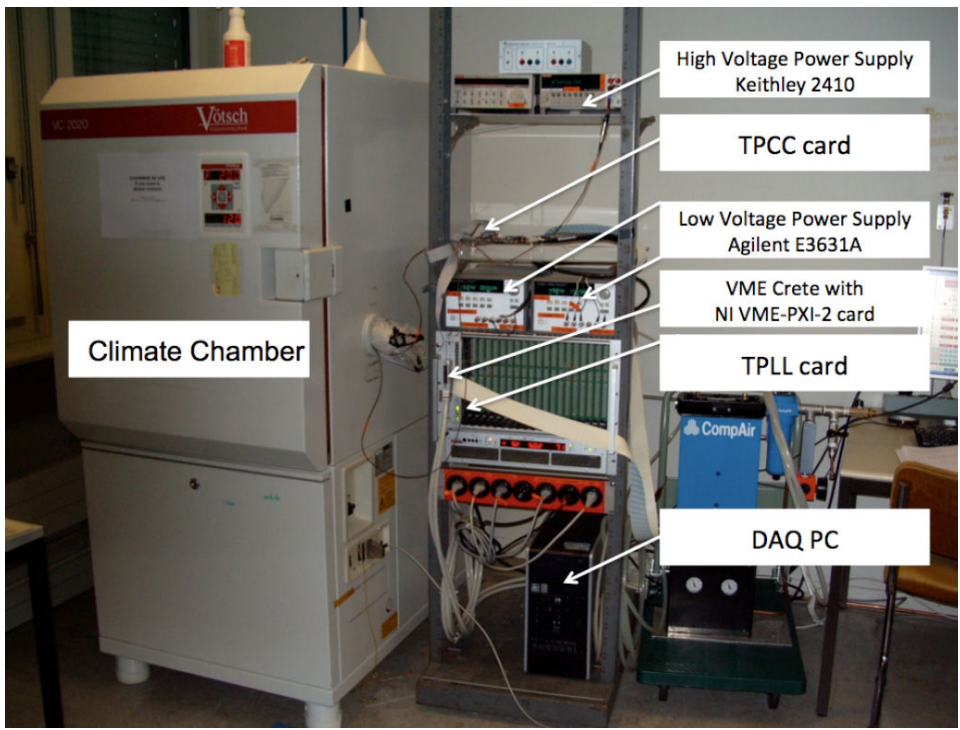

Fig. 4 Snapshot of CERN ATLAS Pixel experimental setup.

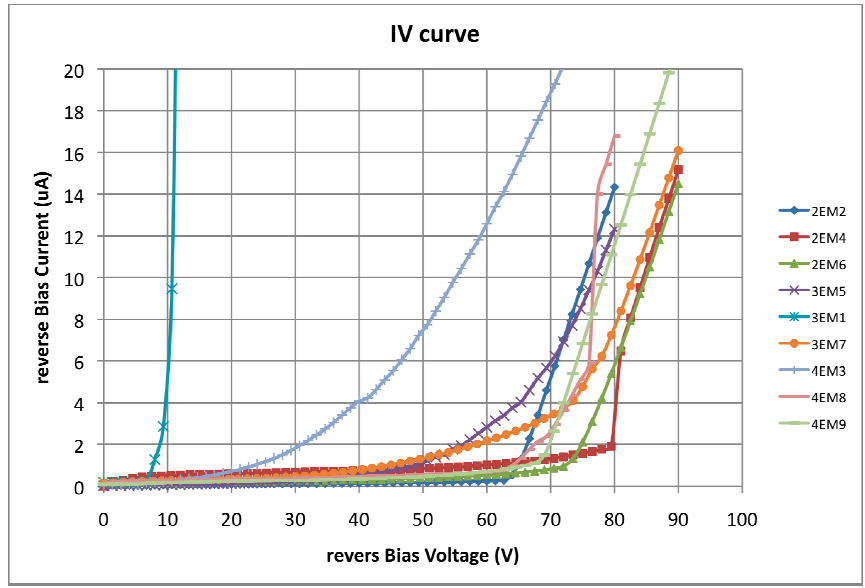

**Fig. 5** Overview of I-V curves of 3D-DDTC sensors from the first batch.

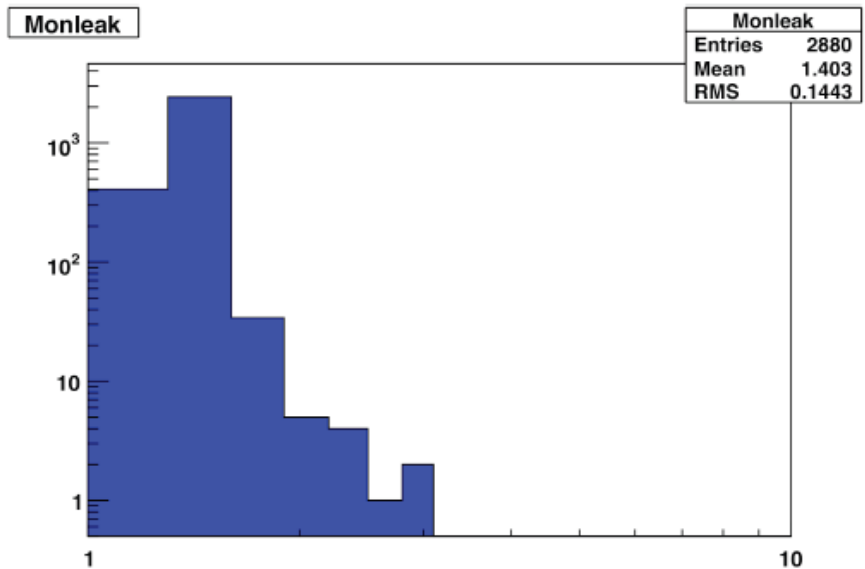

**Fig. 6** Leakage current distribution in DAC units (1 DAC = 125 pA) for a 2E detector module from the first batch biased at 35 V.

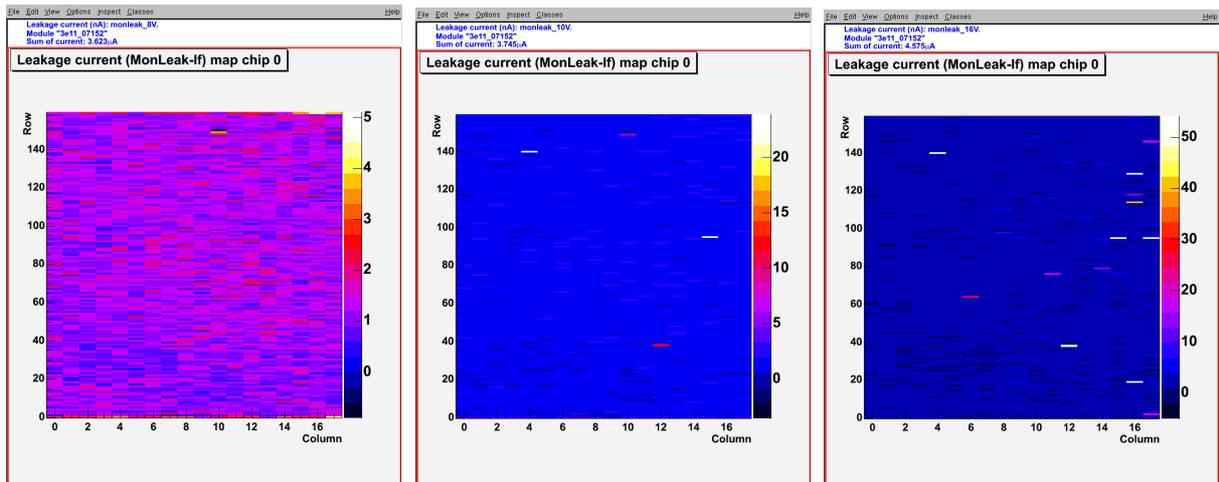

**Fig. 7** Map of leakage current in a 3E sensor from the second batch at different bias voltage: 8V (left), 10 V (centre), and 16 V (right).

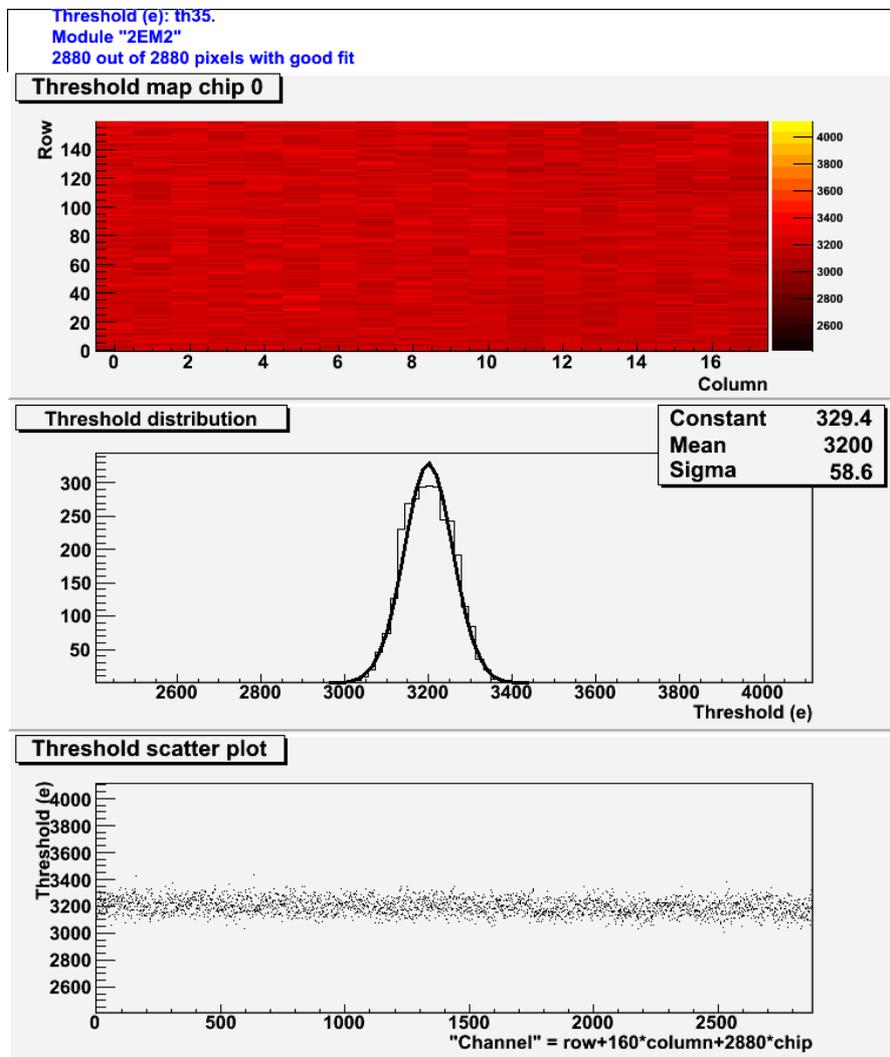

**Fig. 8** Threshold measurement for a 2E sensor from the first batch at a bias voltage of 35 V.

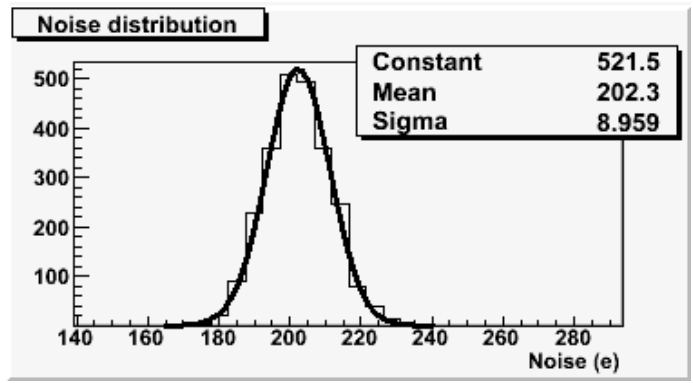

**Fig. 9** Noise distribution for a 2E sensor from the first batch at a bias voltage of 35 V.

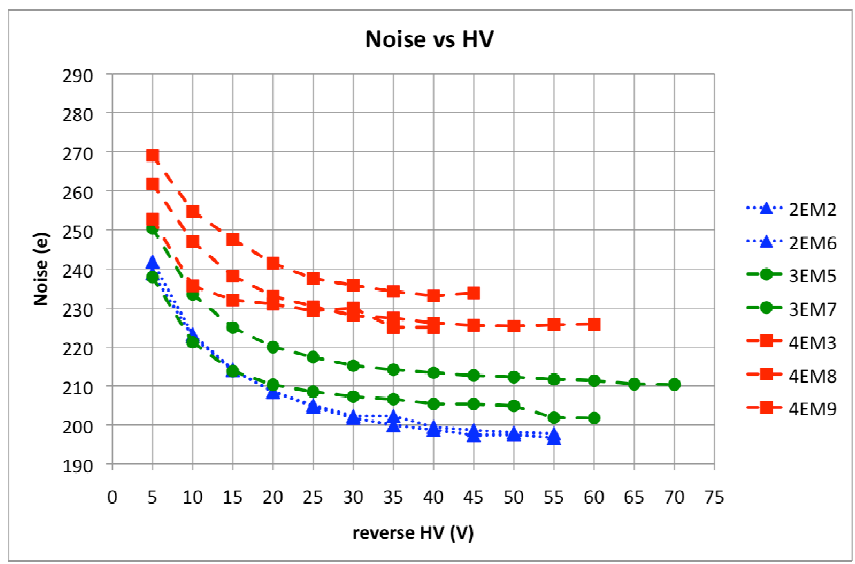

**Fig. 10** Scan of noise versus bias voltage for the "good" sensors from the first batch.

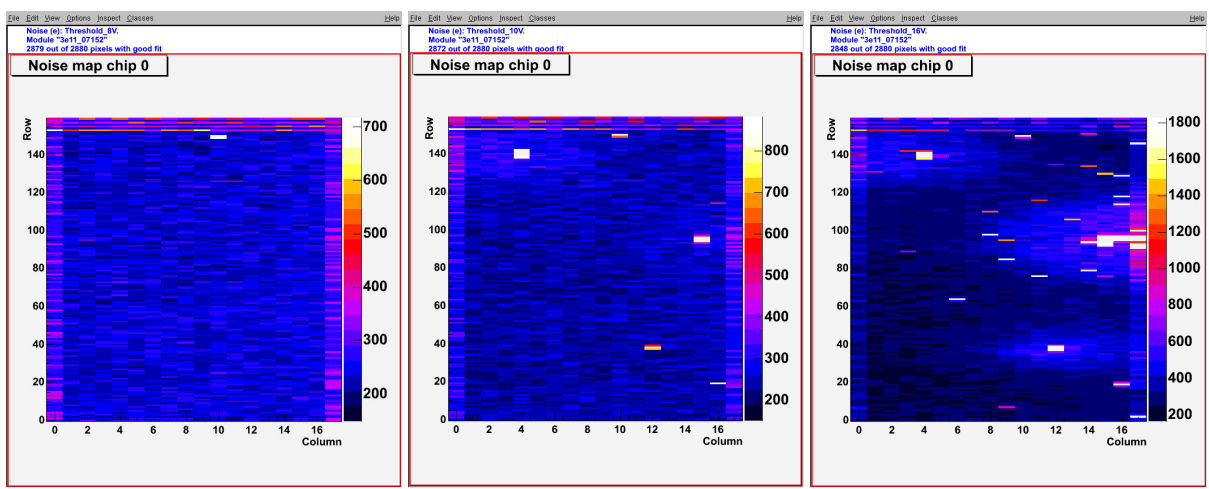

**Fig. 11** Map of noise in the same 3E sensor as in Fig. 9 at different bias voltage: 8V (left), 10 V (centre), and 16 V (right).

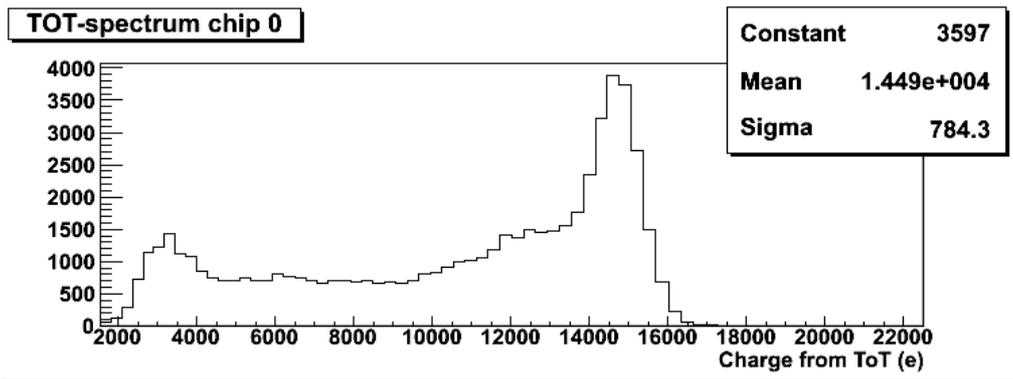

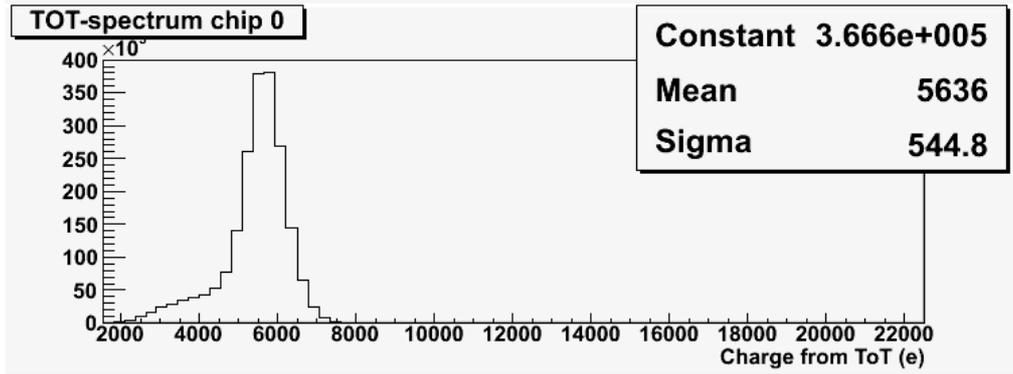

**Fig. 12** Am$^{241}$ (top) and Cd$^{109}$ (bottom) spectra measured with a 2E sensor from the first batch at a 35 V bias voltage. The figures show the source spectrum as a sum over all pixels without any clustering. The reported data refer to a gaussian fit of the main peak (not shown).

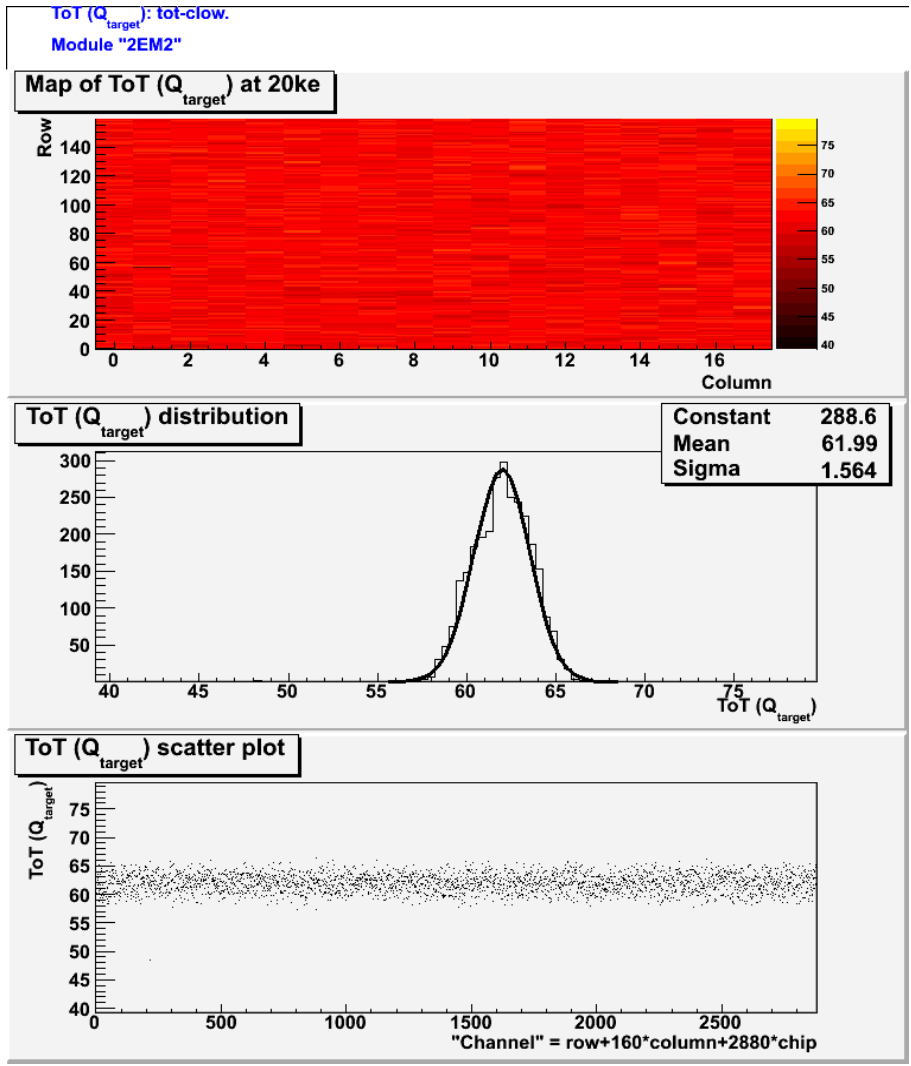

**Fig. 13** ToT calibration for the 2E sensor.

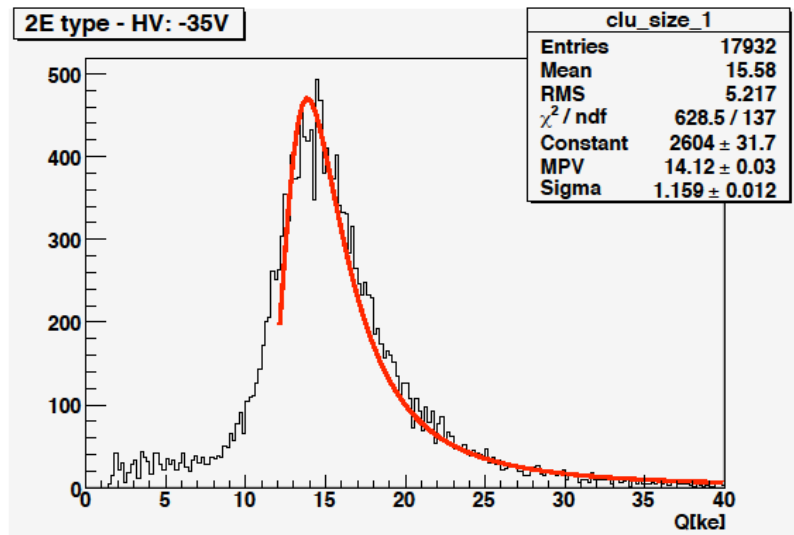

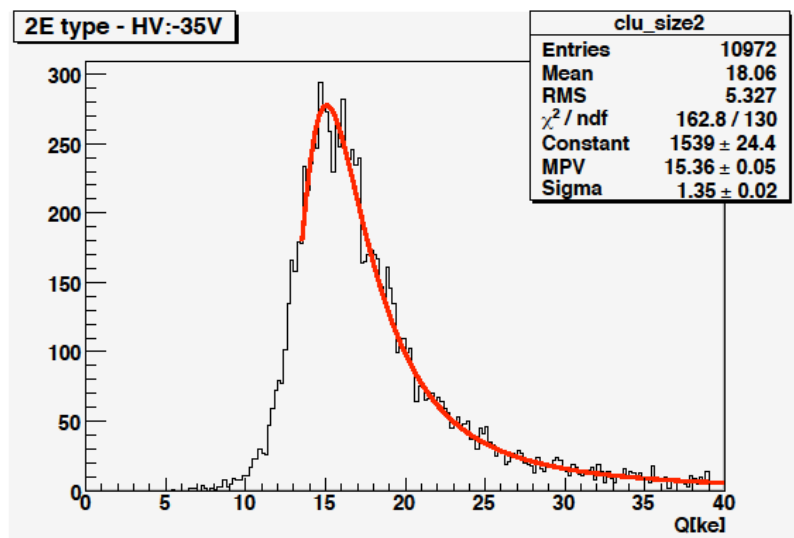

**Fig. 14** Pulse height distributions for cluster size 1 (top) and cluster size 2 (bottom) in a 2E sensor biased at 35 V in response to a Sr$^{90}$ source. The reported data refer to a Landau fit of charge values.